\documentstyle{article} 
\title{\begin{flushright}
{\bf  \normalsize SWAT/86}\\ 
\end{flushright}
\bf Thin Absolute Villains }

\author{ {\it C.F. Baillie} \\
         Computer Science Dept. \\ University of Colorado\\ Boulder, CO 80309,
	 USA\\ \\
         {\it N. Dorey} \\
         Department of Physics\\
         Universite of Wales Swansea\\
         Swansea, SA2 8PP, Wales \\ \\
         {\it W. Janke}\\
         Institut f\"ur Physik\\ 
         Johannes Gutenberg-Universit\"at \\
         D-55099 Mainz, Germany\\ \\
	 and 
         \\ \\ 
         {\it D.A. Johnston}\\
         Dept. of Mathematics\\ 
         Heriot-Watt University\\ 
         Edinburgh, EH14 4AS, Scotland }

\begin{document}
\maketitle
%-----------------------------------------------------------------------
		      {\Large \begin{abstract}
%-----------------------------------------------------------------------
% 
We perform simulations of an
absolute value version of the Villain
model on $\phi^3$ and $\phi^4$ 
Feynman diagrams, ``thin'' 3-regular 
and 4-regular random graphs.
%WJ
The $\phi^4$ results 
are in excellent quantitative agreement with the exact calculations by 
Dorey and Kurzepa
for an annealed ensemble of
thin graphs, in spite of simulating only a {\it single} graph
of each size.
%WJadd
We also derive exact results for an annealed ensemble
of $\phi^3$ graphs
and again find excellent agreement with the numerical data
for single $\phi^3$ graphs.

The simulations confirm the picture of a mean
field vortex transition which is suggested by the analytical results. 
Further simulations 
on $\phi^5$ and $\phi^6$ graphs and of 
the standard XY model on $\phi^3$ graphs 
confirm the universality of these results.
The calculations of Dorey and Kurzepa
were based on reinterpreting the large orders
behaviour of the anharmonic oscillator in
a statistical mechanical context so
we also discuss briefly the interpretation of 
singularities in the large orders behaviour in other models as phase
transitions. 
%
%-----------------------------------------------------------------------
			\end{abstract} }
%-----------------------------------------------------------------------
%
  \thispagestyle{empty} 
%
%***********************************************************************
%
  \newpage 
%
%-----------------------------------------------------------------------
		  \pagenumbering{arabic}
%-----------------------------------------------------------------------

\section{Background and Calculations}

It has long been known that mean field behaviour
is found in models with short range interactions living
on tree-like structures 
such as Bethe lattices \cite{1}. This
approach circumvents
some of the problems that appear 
with using infinite range interactions to
get mean field results.
Difficulties still arise, however, both analytically and numerically
when dealing with the dominant boundary of such trees.
Random graphs, which are locally tree-like and have no external legs,
offer a way round this problem, giving a way of calculating and simulating
models on closed lattices with short range interactions that 
%are 
%WJ
still behave in a mean field like manner.

There has been a considerable amount of work on spin glass models
on random graphs \cite{3} 
mostly with Ising or Potts spins, 
often based
on analogy with the corresponding Bethe lattice.
Recently it was pointed out that transplanting methods
from matrix models 
and 2D quantum gravity allowed a considerable simplification
of many of the proofs that had been derived \cite{4}
and offered the possibility of
attacking problems like replica symmetry breaking
from a different perspective.
In effect one considers a $d=0$, $N=1$ matrix, 
or scalar, model 
where $N$ is the size of the matrix,
to generate the requisite ensemble
of random graphs in a Feynman diagram expansion.
For large enough graphs saddle point methods
can be employed and the mean field critical behaviour calculated
from the algebraic saddle point 
equations\footnote{Inverting
the logic of such an approach to use statistical
mechanical methods to investigate the large orders
behaviour of Feynman diagram expansions was actually
suggested some years
ago by Bender and Wu \cite{4a}.}.

The matrix model inspired approach to discrete spin models
was predated by independent work on 
the finite temperature quantum mechanics of
the anharmonic oscillator, interpreted as
a $d=1$, $N=1$
matrix model \cite{5}. This described
a Villain transcription of the continuous spin $XY$ 
model on $\phi^4$ graphs
and gave a mean field vortex transition
rather than the Kosterlitz-Thouless ($KT$) \cite{6} transition of the 
standard two dimensional $XY$ model.
In essence, the thermodynamic limit of the  
$\phi^4$ random graph model is described 
by large orders in the perturbation series
of the anharmonic oscillator in finite temperature 
quantum mechanics.

Part of the original motivation
for considering the 
XY model transition on an ensemble
of random graphs was the study of the XY
model coupled to 2D quantum gravity. This is 
equivalent to putting the discretized model on an annealed ensemble
of {\it fat}, or planar, graphs. Calculations \cite{7}
and simulations \cite{8} have both indicated
that the transition was still KT in nature in such an ensemble,
but a definitive proof of this has been elusive.
The idea in \cite{5} was
to find an ensemble of graphs in which complete calculations
{\it were} possible and in which some sort of transition
was still manifest. An ensemble of generic ``thin'' random
graphs proved to be such a case. 

If we write the finite temperature partition function
for the anharmonic oscillator as
\begin{equation}
Z ( \beta , g ) = \int D \phi \exp \left( - \int^{\beta}_0
d \tau \left[
\frac{1}{2} \dot \phi^2 + \frac{1}{2} \phi^2 + g \phi^4 \right] \right),
\end{equation}
where $\beta = 1 / T $ is the inverse temperature,
and carry out the perturbative expansion in $g$,
%WJ
\begin{equation}
Z (\beta, g ) \simeq \sum_{k=0}^{\infty} Z_k (\beta) g^k,
\end{equation}
then each $Z_k (\beta)$ may be written
as a sum over Feynman diagrams,
which is this case are random $\phi^4$ graphs with $k$ vertices, giving
\begin{equation}
Z_k (\beta) = (-1)^k \sum_G S(G) \int_0^{\beta} \ldots \int_0^{\beta}
\prod_{i=1}^k d t_i \prod_{<ij>} \sum_{m_{ij}= - \infty}^{\infty}
\exp \left( - | t_i - t_j + m_{ij} \beta | \right).
\end{equation}
In the above $S(G)$ is the symmetry factor, which will generically
be unity for a large graph, the $t_i$ are attached to each vertex
%WJ
and integrated over $0$ to $\beta$, and the $m_{ij}$ are attached to each link
and summed over the integers. The partition function $Z_k (\beta)$
can be thought of as coming from embedding the $\phi^4$ graph
on a circle of period $\beta$. The finite temperature 
one-dimensional propagator that appears in the above,
\begin{equation}
D_{<ij>} = \sum_{m_{ij}= - \infty}^{\infty}
\exp \left( - | t_i - t_j + m_{ij} \beta | \right),
\end{equation}
assigns a time coordinate $t_{i,j}$ to each end of an edge
as well as a winding number $m_{ij}$
to the edge itself. Written in this form the similarity
with the Villain version of the $XY$ model \cite{villain}, where
the edge factor is
\begin{equation}
\tilde{D}_{<ij>} = \sum_{m_{ij}= - \infty}^{\infty} 
\exp \left( - \frac{\beta}{2} (\theta_i - \theta_j +  2 \pi m_{ij})^2 \right),
\end{equation} 
%where $\theta$ has period $2 \pi$,
%WJ
with $\theta$ taking values in the interval 0 to $2 \pi$,
is apparent so we might expect the same sort of critical behaviour.
We have, in effect, interpreted the finite temperature
quantum mechanics of the anharmonic oscillator as an absolute
value version of the Villain model living on thin random graphs.

For convenience here we perform a rescaling of the
$t_i$ to obtain 
\begin{equation}
Z_k (\beta) = (-1)^k \beta^k \sum_G S(G) \int_0^{1} \ldots \int_0^{1}
\prod_{i=1}^k d t_i \prod_{<ij>} \sum_{m_{ij}= - \infty}^{\infty}
\exp \left( - \beta | t_i - t_j + m_{ij} | \right),
\end{equation}
which will simplify some of the later formulae for 
quantities to be measured in the simulations. As in \cite{5}
the free energy per vertex is defined as
\begin{equation}
F = -\lim_{k \rightarrow \infty} { 1 \over k} 
\log | {Z_k ( \beta) \over n_k } |,
\end{equation}
where $n_k$ is the number of graphs of size $k$, $n_k \simeq (16)^k (k
- 1 )!$. Note that, in this case,  
the thermodynamic limit is taken by setting $k \rightarrow \infty$
by hand. Unlike the case of planar diagrams and two dimensional
gravity there is no cosmological constant dual to the area
that one can tune to induce critical behaviour.

The energy per vertex 
in the model $\rho = \partial F / \partial \beta$ measures the
expectation value of the target-space length of the embedded graph, 
as can be seen by differentiating
the Laplace transform of $Z_k ( \beta)$,
%WJ
\begin{equation}
Z_k (\beta) = \int_0^{\infty} dL\, \tilde Z_k (L) \exp ( - L \beta ),
\end{equation}
with respect to $\beta$. As discussed in \cite{5}, $\rho$ can be 
interpreted as a measure of the density of vortices in the model. 
The specific heat is also
given by the standard formula
\begin{equation}
C = - \beta^2 {\partial^2 F \over \partial \beta^2 },
\end{equation}
or equivalently by directly differentiating the 
energy $C = \partial \rho / \partial T$.

The analytical solution of the thin graph model
proceeds by looking at the large orders behaviour
of the anharmonic oscillator partition function 
\cite{9} in equ.(1).
%WJadd
After a rescaling $x = 2\sqrt{-g}\phi$,
the $k \rightarrow \infty$ saddle point solution
is given by the trajectories of period $\beta$ of a 
%WJ check: is this already rescaled ?????
particle moving in the potential $V = -\frac{1}{2}x^2 + \frac{1}{4}x^4$.
One uses the dispersion relation,
%WJ
\begin{equation}
Z_k ( \beta ) = \frac{1}{\pi} \int_{- \infty}^0 dg 
{ {\rm Im} Z(\beta, g + i 0) \over g^{k+1} },
\end{equation}
along with the saddle-point evaluation of ${\rm Im} Z(\beta,g)$ 
to extract $Z_k (\beta)$. A non-trivial instanton solution only exists
above a critical value of the period $\beta$ 
(in the $\phi^4$ case $\beta_{c}=\sqrt{2} \pi$). For $\beta<\beta_{c}$
the only contribution is from the
trivial solution $x(t)\equiv x_{\rm min}$, where the particle sits at the
bottom of the well. The change of behaviour at $\beta=\beta_{c}$ is taken
as the signal for a phase transition in the 
associated absolute value Villain model.
In the low temperature phase we find 
\begin{equation}
Z_k \simeq (-1)^k {\beta \over 2} \left( {1 \over 2 \pi} 
{\partial E \over \partial \beta } \right)^{1/2} 
\left[ I(\beta)/4 \right]^{-(k+1/2)} \Gamma ( k+ 1/2),  %WJ
\end{equation}
while for $\beta<\beta_{c}$, 
\begin{equation}
Z_k \simeq (-1)^k 2 \left( { \sqrt{2} \sinh (\beta) \over
\sinh ( \sqrt{2} \beta) } \right)^{1/2} \left( {\beta \over 16}\right)^{-k}
\Gamma ( k ).
\end{equation}
In the above $I( \beta )$ is just the (scaled) classical action 
%WJ
%associated with a trajectory and
of the instanton satisfying
\begin{equation}
{\partial I ( \beta ) \over \partial \beta } = - E ( \beta), %WJ
\label{eq:I_E}
\end{equation}
where $E ( \beta)$ is
%the energy at a turning point of the motion. 
the energy associated with this trajectory.
To obtain 
explicit expressions in the above we need to determine the
dependence of $E$ on the period $\beta$. This can be done by
evaluating the first integral of the classical equations of motion, 
\begin{equation}
\beta (E ) = 2 \int_{x_1}^{x_2} {  dx \over \sqrt{ 2 [ E - V ( x ) ] }},
\end{equation}
%WJ
where $x_1$ and $x_2$ are the turning points.
In the $\phi^{4}$ case, 
the above integral can be evaluated perturbatively near $\beta_{c}$
\cite{5}. Inverting the power series gives 
$E (\beta) = - 1 / 4 + 4 ( \beta - \beta_c) / (3 
\beta_c) + \ldots$. 
%WJ check: unclear
These solutions predict $\rho = 1 / \beta$ for $\beta<\beta_c$
and $\rho \simeq \exp ( - \beta ) $ at large $\beta$. Similarly
$C = 1$ for $\beta < \beta_c$ and sweeps up to a sharp cusp
of height $19/3$ as $\beta \rightarrow \beta_c+$.
%WJadd
In order to obtain the full temperature dependence of $\rho$ and $C$ 
shown in Figs.~1 and 2, we found it useful to express both $\beta(E)$ and the 
(scaled) classical action in terms of elliptic integrals, which can easily
be evaluated with high precision. 
In summary, we see a second order transition of the mean-field
type, with a sharp cusp
discontinuity in the specific heat.

An explicit calculation can also be carried through in 
a similar style for the $\phi^3$ case
where the partition function is 
\begin{equation}
Z ( \beta , g ) = \int D \phi \exp \left( - \int^{\beta}_0
d \tau \left[
\frac{1}{2} \dot \phi^2 + \frac{1}{2} \phi^2 + g \phi^3 \right] \right).
\end{equation}
After a rescaling $x = -3 g \phi$, the problem is
mapped onto considering the trajectories of a particle
in the potential $= -\frac{1}{2}x^2 + \frac{1}{3} x^3$.
The relation (\ref{eq:I_E}) between the classical action %$I(\beta)$   %WJ
and the 
%WJ check: what is "turning point" energy ????
%turning point 
energy 
%is now
%\begin{equation}
%{\partial I ( \beta ) \over \partial \beta } = - \frac{1}{9} E ( \beta),
%\end{equation}
%which, as
%WJ
%for the $\phi^4$ potential, 
%in the $\phi^4$ case, 
can be solved, as in the $\phi^4$ case, by inverting the relation
for $\beta(E)$
to get $E(\beta)$. Here we find
\begin{equation}
\beta =  6^{1/3} { 2 K (k) \over \sqrt{x_3 - x_1}}, %WJ 1 <-> 3 check!!!
\end{equation}
where $K$ is the complete elliptic integral with modulus
$k = \sqrt{ ( x_3 - x_2) / (x_3 - x_1)}$,
%WJ
%and the arguments $e_{1,2,3}$ are determined from the 
%turning points of the motion as the roots 
and $x_1<x_2<x_3$ are the roots of
%of the equation 
$ 2 [ E  - V(x) ] = 2E + x^2 - 2 x^3 / 3 =0$.
Inverting gives $E (\beta) = - 1 /6 + 6 (\beta - \beta_c) / ( 5 \beta_c)
+ \ldots$,
where $\beta_c$ is now $2 \pi$.
The form of $\rho$ and $C$ is similar
to the $\phi^4$ graphs, but we now find a peak
of $41/10$ in $C$ as $\beta \rightarrow \beta_c+$
and $C = 1/2$ for $\beta < \beta_c$.
As in the $\phi^4$ case, to compute the full temperature dependence of 
$\rho$ and $C$,
we first expressed the (scaled) classical action
in terms of elliptic integrals, 
%$I = (4/15)\sqrt{2/3}(x_3-x_1)^{5/2}\left[
%(2-2k^2+2k^4)E(k) - (2-3k^2+k^4)K(k)\right] - E\beta$.
\begin{equation}
I = {4 \over 15} \sqrt{2 \over 3}(x_3-x_1)^{5/2} \left[
(2-2k^2+2k^4)E(k) - (2-3k^2+k^4)K(k) \right] - E\beta.
\end{equation}

A couple of other numerical consequences of the 
analytical 
results are worth remarking upon as they are convenient for
verifying that simulations are performing correctly.
Firstly, we find that $\rho (T_c)$ is equal to $T_c$
on $\phi^4$ graphs.
Secondly, it should be noted that the $\beta_c$ is determined
only by the period of oscillations around the minimum of
the potential for {\it any} potential. 
Approximating this region with a quadratic
gives 
$\beta_c = 2\pi / m$ if  %WJ normaliztaion of m changed
$V \simeq - V_0 + m^2 ( x - x_0)^2/2$, where $x_0$ is the minimum
point.
This reproduces the explicit results
derived above for $\phi^3$ and $\phi^4$ graphs with rather less
pain. 
Finally, it is worth noting that, for a potential
with an anharmonic term
of the form $\phi^{2(p+1)}$, we will have $C =  p$ for $T \ge
T_{c}$ because  
$\rho = p T$ for $T \ge T_c$ in general.

We can contrast the critical behaviour 
described above with 
the standard $XY$ model
on a flat two dimensional lattice
%WJ
\begin{equation}
Z= \prod_i \left[ \int_0^{2\pi}\!\!\! d \theta_i \right] 
\exp \left( \beta \sum_{<ij>} 
\cos ( \theta_i - \theta_j) \right),
\end{equation}
which displays a 
topologically driven Kosterlitz-Thouless (KT) transition.
The specific heat has only a broad cusp rather
than a divergence.
However, the
correlation length has a critical singularity
\begin{equation}
\xi = A_\xi \exp \left( { B_\xi
\over (T - T_c)^\nu } \right),
\end{equation}
as does the spin
susceptibility
\begin{equation}
\chi = A_\chi \exp \left( { B_\chi
\over (T - T_c)^\nu } \right).
\end{equation}
The
exponent $\nu$ is predicted to be $1/2$ in the KT theory and
the correlation function critical exponent is
predicted to be $\eta = 1/4$, where $\eta$
is given by
%WJ
\begin{equation}
\chi \propto \xi^{2-\eta}.
\end{equation}
In spite of the differences the physical picture of
the transition on thin graphs is still very similar 
to that of the standard KT transition.
Given the interpretation of $\rho$ as a vortex density,
the preceding saddle point results 
show that as the temperature is increased
(ie $\beta$ is decreased) vortices are liberated, with the
vortex density increasing by almost an order of magnitude
around $\beta_c$.
 
As we have already noted, it is not a
foregone conclusion that putting the model on {\it any} collection
of random graphs will give mean field behaviour - the
model still displays a KT transition on an annealed
set of planar random graphs (ie when coupled to 2D
quantum gravity). An explicit check of the thin graph predictions
is therefore not totally vacuous.
In what follows we 
describe some modest simulations 
that we carried out in order to verify the
mean field nature of the transition
for variants of the model on various
thin graphs. 
We are essentially interested
in the behaviour of the energy and specific heat with
$\beta$ and determining $\beta_c$.
We simulated the absolute value Villain action
on $\phi^3, \phi^4, \phi^5$ and $\phi^6$ random graphs, as well
as looking at the standard $XY$ action of equ.(18)
on $\phi^3$ graphs
in order to check it had the same critical behaviour.

\section{Simulations}

We need to exercise a little care in defining our
observables in the simulation because of the unusual
form of the Boltzmann factors in the partition function
\cite{10}.
If we define the auxiliary sums
\begin{eqnarray}
\Sigma_0 &=& \sum_{m_{ij}= - \infty}^{\infty}
\exp \left( - \beta | t_i - t_j + m_{ij} | \right), \nonumber \\
\Sigma_1 &=& \sum_{m_{ij}= - \infty}^{\infty}
| t_i - t_j + m_{ij} | \exp \left( - \beta | t_i - t_j + m_{ij} | \right), 
\\
\Sigma_2 &=& \sum_{m_{ij}= - \infty}^{\infty} 
| t_i - t_j + m_{ij} |^2 \exp \left( - \beta | t_i - t_j + m_{ij} | \right),
\nonumber
\end{eqnarray}
then the definition of the energy 
$\rho = \partial F / \partial \beta$
applied to the partition function
in equ.(6) gives
\begin{equation} 
\rho = \frac{1}{k} \left< \sum_{<ij>} { \Sigma_1 \over \Sigma_0 } \right>
- \frac{1}{\beta}
\end{equation}
for the energy per site, where the $< \; >$ denote
a thermal average and the additional $1 / \beta$ comes from the
overall factor of  
$\beta^k$ that appears when the $t_i$ are rescaled.

The specific heat can be obtained in the simulations either
by direct numerical differentiation of the measurements of $\rho$
using $C = \partial \rho / \partial T$, or by differentiating
equ.(6) twice to give
\begin{equation}
C = \beta^2 k \left( <\rho^2> -<\rho>^2 \right) + \frac{\beta^2} {k}
\left< \sum_{<ij>} \left( {\Sigma_2 \over \Sigma_0} - { \Sigma_1^2 \over
\Sigma_0^2 } \right) \right>  - 1.
\end{equation}
The second term is non-canonical and is due to
the summed Boltzmann factors whereas the $-1$ results from the
overall $\beta^k$.

Having decided on our observables, it now remains to 
choose an update scheme for the simulation. A simple 
Metropolis update can be used quite efficiently by
adopting a discrete approximation to the periodic
$t_i$ and then tabling the resulting Boltzmann factors
and associated sums $\Sigma_{0,1,2}$, which can then be
looked up during the course of the simulation \cite{10}.
%WJadd
Depending on the temperature, we took from $100$ to $1000$ 
different $t$ values for the tables,
which were constructed by truncating the sum over $m_{ij}$
at $\pm 100$. Increasing these limits made no appreciable difference
to the measured results in any of the simulations reported here.
In doing this we are taking a $Z_{100} \dots Z_{1000}$ approximation
to the $O(2)$ symmetry of the model. Notice that for the absolute value
version of the Villain model one has to be quite careful with this
approximation, in particular at low
temperatures, since the associated discretization error enters
linearly in the action and not squared as in the standard Villain model.
It is perhaps worth
remarking that one could equally well envisage 
leaving the $m_{ij}$ as free variables
on each link to be sampled in the course of the simulation,
but previous work in which the sum is carried out {\it a priori}
as here has given good results for the standard Villain/XY model
\cite{10} and we stick to this.
The direct form of $D_{<ij>}$ converges rapidly for large
$\beta$, but is also possible to use a ``dual''
representation that is best suited for small $\beta$
simulations,
\begin{equation}
\tilde D_{<ij>} = \frac{2}{\beta} \left( 1 + \sum_{k_{ij} = 1}^{\infty}
{ \cos ( 2 \pi k_{ij} ( t_i - t_j)) \over 1 + \left({2 \pi k_{ij} \over \beta} 
\right)^2 } \right).
\end{equation}
The numerical differences, even when using the representations
well into the ``wrong'' regions of $\beta$ for both, are slight.

The final ingredient in the simulations is the choice of
a random graph. The calculations we have outlined in the first
section are supposedly for an annealed ensemble of thin 
graphs, so in theory we should carry out ``flip'' moves
in the same fashion as in simulations of 2D gravity on planar
graphs. However, we can evade this responsibility
by appealing to previous simulations of the Ising model
on random graphs \cite{4}, where a {\it single} random graph of
a given size was enough to extract the mean field
critical behaviour predicted in an annealed calculation.
As the thin random graphs are essentially tree-like
it appears that one tree is very much like another
as far as a ferromagnetic transition is concerned, although
a (quenched) average over graphs becomes essential when
non-self-averaging transitions such as those in spin
glasses are simulated.
We shall take the lazy man's approach here and verify
that the single graph results are consistent with the
analytical calculations that, strictly speaking, only apply
to an annealed ensemble.

We simulated 
the absolute value Villain
model of equ.(6) on $\phi^3$ and
$\phi^4$ graphs of size up to $2500$ vertices,
as well as small runs of $N=250$ $\phi^5$ 
and $\phi^6$ graphs.
In all the cases we carried out 500,000 Metropolis
sweeps at each $\beta$ value, with a measurement
every tenth sweep, after allowing a suitable
amount of equilibration time. As we have indicated
{\it no} flip moves were carried out on the 
graphs concerned. The energy and specific heat,
defined as above, were the principal observables.
In addition, we also simulated the standard
XY model of equ.(18) on $\phi^3$ graphs,
using a single cluster update, largely
as a check on the universality
of the Villain model results.

Turning now to the results themselves we can see
in Fig.~1 that the energy matches closely the predictions
of \cite{5}. There is a low temperature exponential
growth with a ``knee'' at the phase transition
$T = 1 / \sqrt{2} \pi$ ($\simeq 0.225$) followed
by linear growth at larger $T$. It is also
%WJ
clear from Fig.~1 that $\rho (T_c) = T_c$, as
predicted by the analysis in the previous section.
We can obtain the specific heat either from direct
differentiation of the energy or by measuring
the observable defined in equ.(24), and both are plotted
in Fig.~2
for a graph of size $1000$. 
This is the smallest graph size at which the 
peak reaches its expected value ($19/3$) - larger graphs give similar
results, whereas the peak is appreciable lower and more
rounded on the smaller graphs simulated. Surprisingly, it is the
numerical differentiation that gives the closest fit to
the analytical results, perhaps an indication that our statistics
around the phase transition point are rather too modest
for complete numerical accuracy in quantities involving
variances, such as the directly measured specific heat.

Fortified by the good agreement between the analytical
and numerical results for $\phi^4$ graphs 
we can move on to look at the possible variations on the theme
that were outlined in the introduction. We 
consider the $\phi^3$ graphs first. In Fig.~3 we plot
the energy for various graph sizes, showing clearly the
similarity with the $\phi^4$ results. The ``knee''
in the curve is at the expected value of $T_c = 1 / 2 \pi
\simeq 0.16$
\footnote{In this case, however, we do
not expect that $\rho (T_c)=T_c$, but rather 
$\rho (T_c)=T_c/2$.}. As our data points are rather scarcer
than for the $\phi^4$ graphs, the specific heat
peak obtained by numerical differentiation 
as shown in Fig.~4 is not so convincing,
but it is clearly of the same general form and
has the correct large $T$ limit of $1 / 2$.

We have not carried out such extensive simulations
of the $\phi^5$ and $\phi^6$ random graphs, simply
contenting ourselves with verifying that the general
form of the energy is similar and that the large $T$
limit is correct. In Fig.~5 the energy is plotted
up to very large $T$ for $\phi^3, \phi^4, \phi^5$
and $\phi^6$ graphs of size $250$. From the slopes
it is clear that the specific heat prediction
$C \rightarrow p$ for $T \rightarrow \infty$
on $\phi^{2(p+1)}$ graphs is satisfied to
a high degree of accuracy. The finite size effects
for a given graph size increase with the degree
of the vertices, which agrees with the
intuitive picture of $\phi^{2(p+1)}$
graphs being more ``tree-like'' for smaller
$p$ with a given number of vertices.

All of the simulations and discussions so far have dealt with the
absolute value Villain model on various random graphs.
We have done this because the edge
factor in this case is just  the
one dimensional finite temperature propagator of equ.(4),
which allows us
to borrow (steal!) the results from large orders expansion 
of various anharmonic oscillators. We would expect that
the standard $XY$ model of equ.(14) would still give
us similar results on grounds of universality, but as we have
no analytical calculations to fall back on in this case
it is worthwhile verifying this explicitly with simulations.
We therefore simulated the standard $XY$ model
on $\phi^3$ graphs of various sizes, with similar
statistics to the Villain model simulations but
using a single cluster update for improved efficiency.

%WJ check: I don't like this sentence at all
%The energy in the standard $XY$ model does not have a simple
%vortex density interpretation, so we do not plot it here.
The specific heat,
measured directly in the simulation,
is plotted in Fig.~6 for various lattice sizes,
where it is clear that, although the small and large $T$ limits
are different from the Villain models ($1 / 2$ and $0$ respectively),
there is still a sharp cusp in the curve. 
This would again indicate
a transition of mean field rather than $KT$ type,
where there is a much gentler bump in the specific heat curve
away from the phase transition point.

%WJadd
As for the standard Villain model \cite{mapping}, the differences can be 
understood by an approximate mapping of the absolute value version of the
Villain model onto the cosine model. By adapting the formulas in 
\cite{mapping} to the present case, we find that the temperature scales 
should be related by $I_1(\beta^{\rm cos})/I_0(\beta^{\rm cos}) = 
(\beta/2\pi)^2/\left(1 + (\beta/2\pi)^2\right)$, where $I_{0,1}$ are 
Bessel functions and $\beta^{\rm cos}$ denotes the inverse temperature
of the cosine model. Inserting $\beta_c = 2\pi$ this predicts 
$I_1(\beta_c^{\rm cos})/I_0(\beta_c^{\rm cos}) = 1/2$ or
$T_c^{\rm cos} = 0.8625\dots$, in good agreement with the peak
location observed in Fig.~6.

\section{Conclusions and Other Models}

The saddle point predictions for the 
energy and specific heat of the absolute value
Villain model on 
various random graphs are verified by the simulations we have
carried out. As one might expect the standard XY model 
%WJ
on ``thin'' $\phi^3$ graphs behaves
in an analogous fashion, with a mean-field-like transition rather
than a KT transition. 
The second point worth emphasizing is that we have not needed
to simulate an annealed ensemble of random graphs
to get good {\it quantitative} agreement with the theory, just as
for the Ising ferromagnet on thin graphs \cite{4}. This should be
contrasted with the planar graphs in 2D gravity where an annealed
sum, usually implemented by flip moves in a simulation, appears
to be essential. It appears that, unless one is very unlucky,
any large thin graph is as good as another. Trees, and the near trees
that thin random graphs represent, all really do look much alike.

The methods used in deriving the analytical results are
borrowings from standard large orders lore
in quantum mechanics. It is rather
the interpretation
in terms of a statistical mechanical model that is novel,
just as the behaviour of discrete spin
models on random graphs is extracted from 
known saddle points results for ordinary integrals in \cite{4}.
We are not restricted
to the simple anharmonic oscillator
in searching for statistical
mechanical interpretations
of large orders behaviour in quantum mechanics. 
Another possible example was considered by
one of the authors of this paper in \cite{11}, namely
the quantum mechanics of an anisotropic
anharmonic oscillator, where the partition function is
%WJ
\begin{equation}
%Z = \int D x D y \exp \left( - \int_0^{\beta} 
Z = \int D \phi_1 D \phi_2 \exp \left( - \int_0^{\beta} 
d \tau \left[
%\frac{1}{2} ( \dot \phi_1x^2
\frac{1}{2} ( \dot \phi_1^2
%+ \dot y^2 + x^2 + y^2) + g ( x^4 + 2 c x^2 y^2 +y^4 ) \right] \right).
+ \dot \phi_2^2 + \phi_1^2 + \phi_2^2) + g ( \phi_1^4 + 2 c \phi_1^2 \phi_2^2
+\phi_2^4 ) \right] \right).
\end{equation}
Taking $c$ as the control parameter rather than $\beta$, the large orders
behaviour shows a transition at $c=1$, where the model is
rotationally symmetric. For $-1 \le c <1$
%the quartic $x^4+y^4$ terms dominate and the instanton
the quartic $\phi_1^4+\phi_2^4$ terms dominate and the instanton
%solution is $x(t) = u(t), y(t)=0$ where
solution is $\phi_1(t) = u(t), \phi_2(t)=0$, where
\begin{equation}
u(t) =  \sqrt{ 1 \over 2 | \lambda | } {1 \over \cosh ( t - t_0)},
\end{equation}
%whereas for $c>1$ the $x^2 y^2$ term dominates and the solution
whereas for $c>1$ the $\phi_1^2 \phi_2^2$ term dominates and the solution
%is of the form $x(t) = y(t) = u(t) / \sqrt{2}$.
is of the form $\phi_1(t) = \phi_2(t) = u(t) / \sqrt{2}$.
Looking at the Feynman diagrams generated by the model
%we can see that it is a sort of loop gas, with $x$ loops and $y$ loops
we can see that it is a sort of loop gas, with $\phi_1$ loops and $\phi_2$ loops
%mixing via the $x^2 y^2$ vertex and a propagator $D_{<ij>}$
mixing via the $\phi_1^2 \phi_2^2$ vertex and a propagator $D_{<ij>}$
between the individual vertices on loops of both types.
A matrix model style order parameter can be defined as
%WJ check: where to put expectation values < > ????
\begin{equation}
%M = { x^4 - y^4 \over x^4 + y^4}
M = { \phi_1^4 - \phi_2^4 \over \phi_1^4 + \phi_2^4},
\end{equation}
which, using the instanton solutions, gives $M=1$ for $-1 \le c <1$
and $M=0$ for $c>1$. 
Alternatively, we could consider the ratio of mixed
to pure vertices,
\begin{equation}
%\tilde M =  1 - { 2 x^2 y^2 \over x^4 + y^4}
\tilde M =  1 - { 2 \phi_1^2 \phi_2^2 \over \phi_1^4 + \phi_2^4},
\end{equation}
with similar results.
With either order parameter
we can see that the random graphs are filled with
$x$ loops only for $-1 \le c <1$ and an even mixture
of $x$ and $y$ loops for $c>1$. 
The singular behaviour at  $c=1$ can thus
be viewed as a sort
of magnetization transition.

It would be an interesting exercise to see if other
large orders results in quantum mechanics
could be recast in a statistical
mechanical mould as in this case. We leave this for
future work.

\section{Acknowledgements} 
CFB is supported by DOE under
contract DE-FG02-91ER40672 and by NSF Grand Challenge Applications
Group Grant ASC-9217394. ND is supported
by a PPARC advanced fellowship. WJ
thanks the Deutsche Forschungsgemeinschaft for a Heisenberg fellowship.
This work has been carried out in the framework of the EC HCM network
grant ERB-CHRX-CT930343.

\vfill \eject 
\bigskip

\centerline{\bf Figure Captions} 

\begin{description}
\item[Fig. 1.] The energy for various $\phi^4$ graph sizes.
A dotted line indicates the analytical prediction.
\item[Fig. 2.] The specific heat for a $\phi^4$ graph of size
$1000$,
obtained via both numerical differentiation 
of $\rho$ and direct measurement. Again the dotted line
represents the analytical curve.
\item[Fig. 3.] The energy for various $\phi^3$ graph sizes.
The dotted line indicates the analytical prediction.
\item[Fig. 4.] The specific heat for  $\phi^3$ graphs of 
various sizes, obtained via numerical differentiation of $\rho$.
%WJ
The dotted line shows the analytical prediction.
\item[Fig. 5.] The energy for $\phi^3$, $\phi^4$, $\phi^5$ and $\phi^6$
graphs. The linear prediction, $\rho = p T$ on $\phi^{2(p+1)}$
graphs when $T \ge T_c$, is shown as dotted lines
to emphasize the very good fit.
\item[Fig. 6.] The specific heat for the {\it standard} XY model
on $\phi^3$ graphs of various sizes.
\end{description}

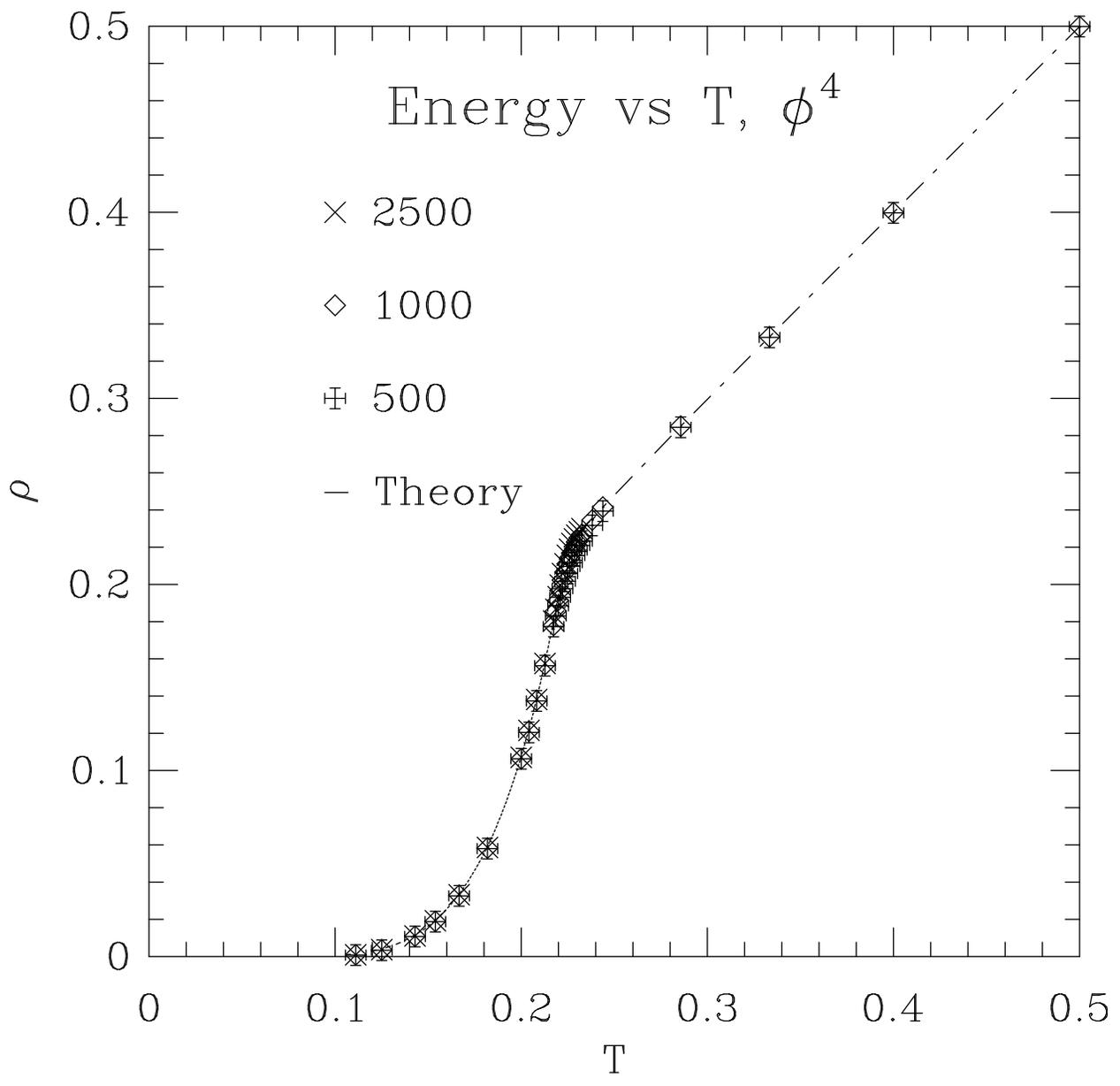
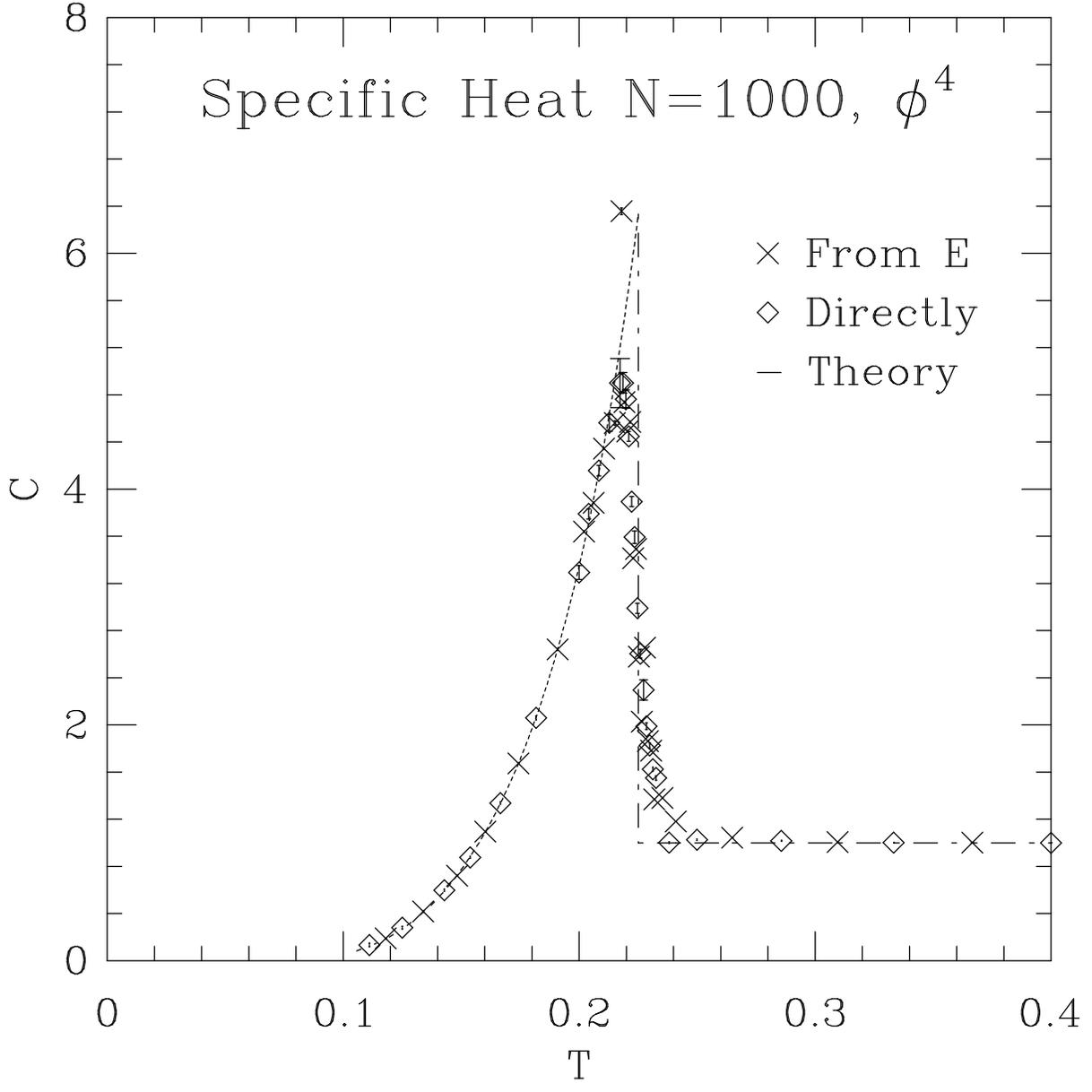
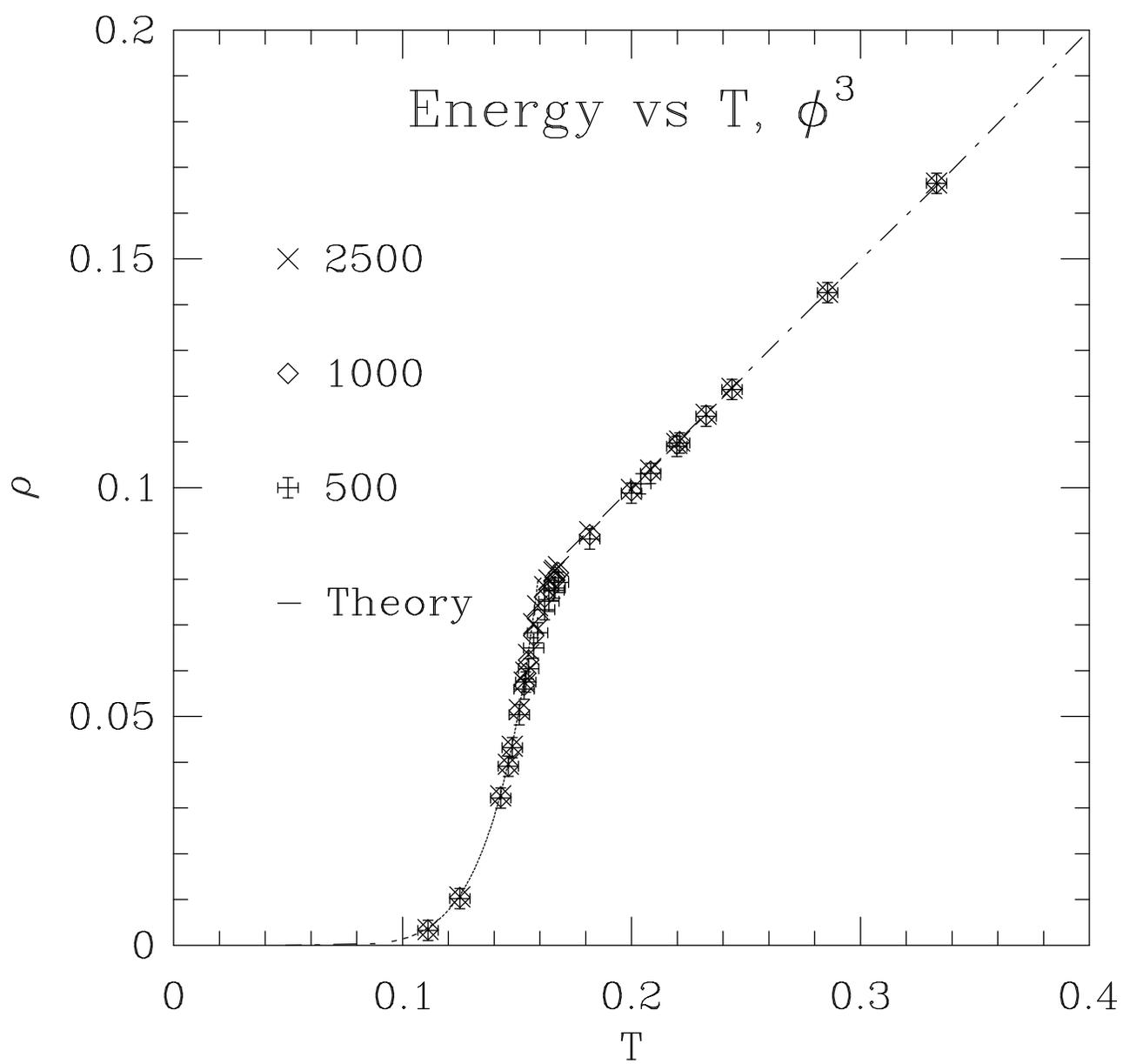
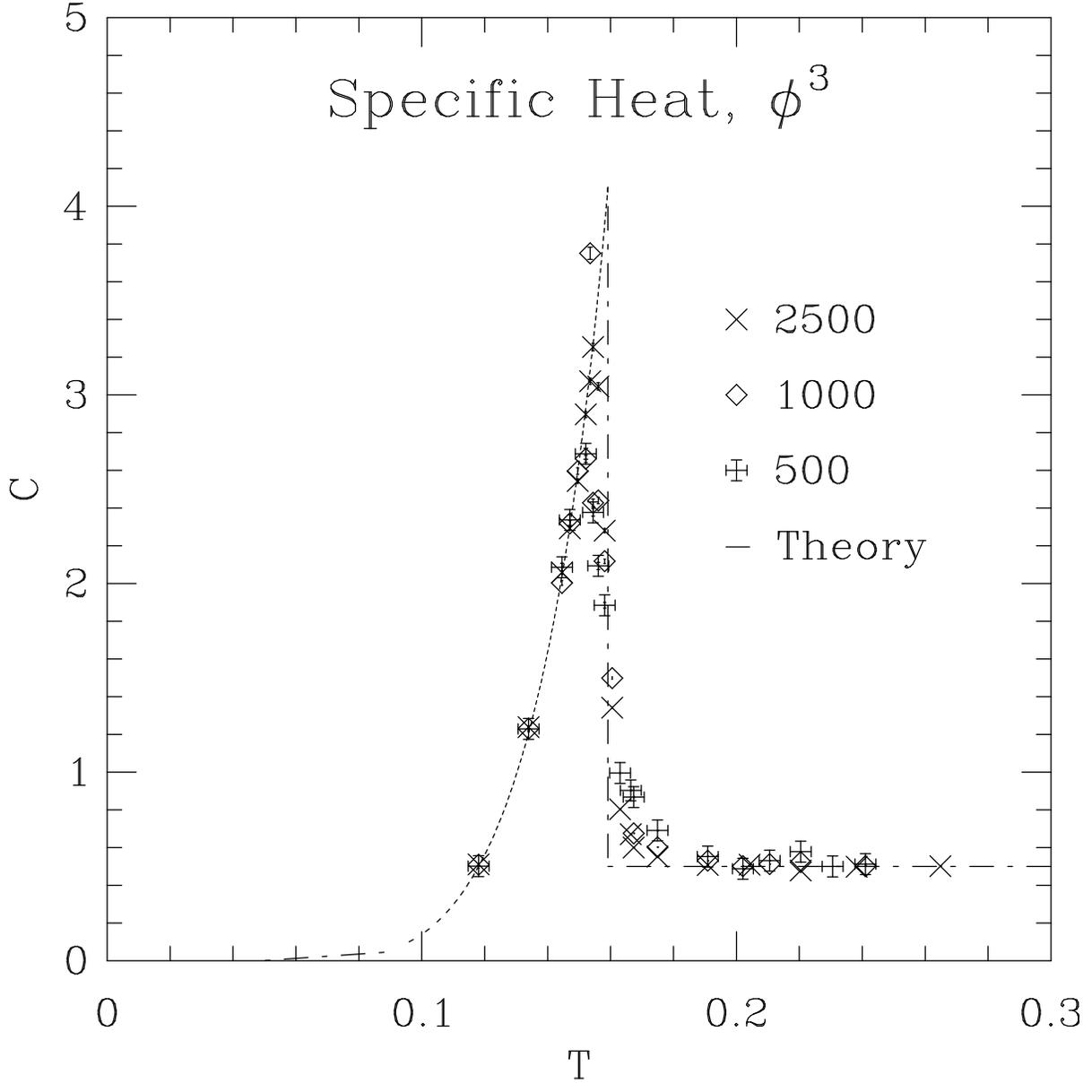
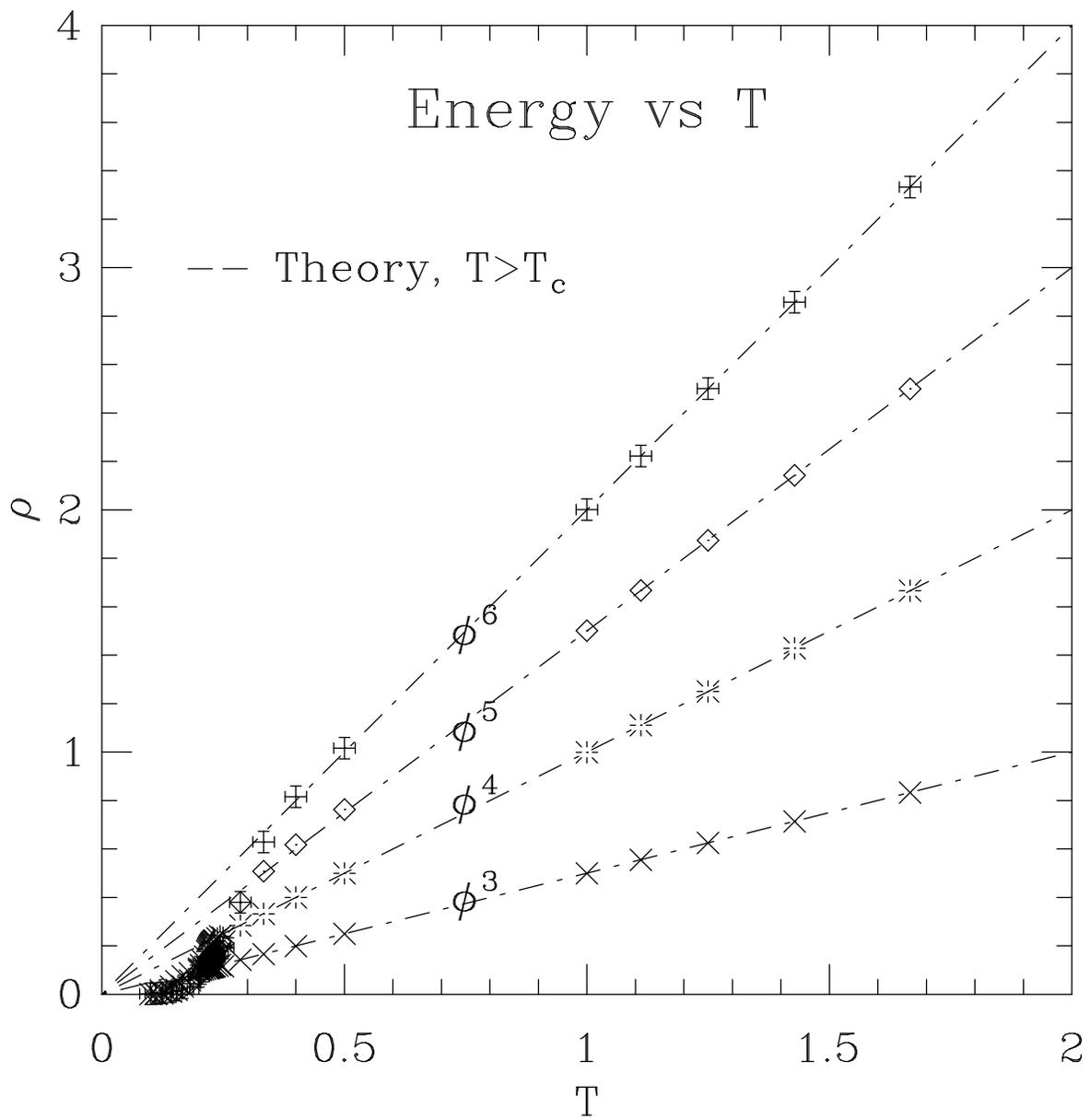
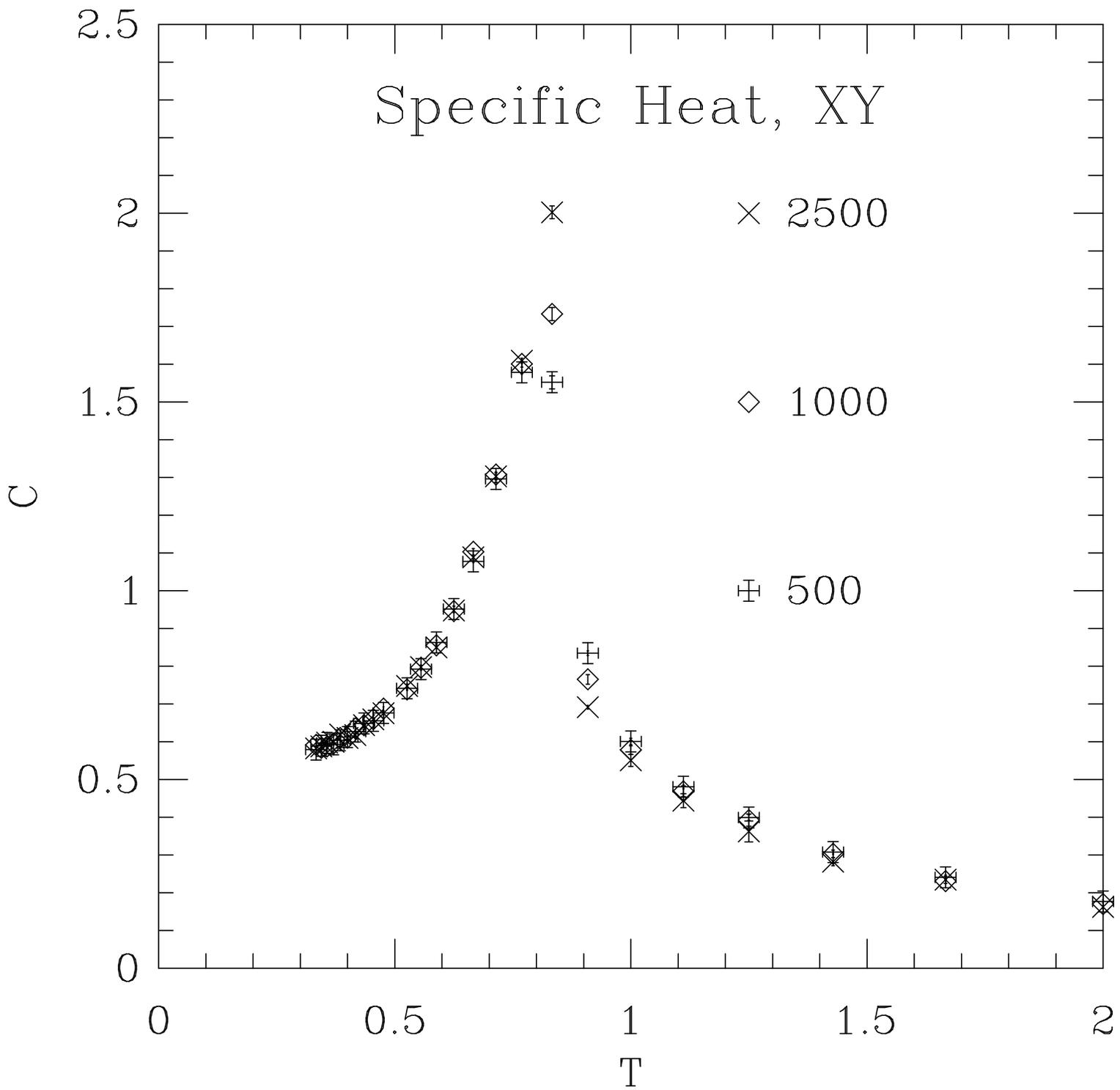
\end{document}